\documentclass[doublecol]{epl2} 

\newcommand{\pp}[2]{\frac{\partial {#1}}{\partial {#2}}}
\newcommand{\ten}[1]{\stackrel{\leftrightarrow}{\bi{#1}}}

\newcommand{\di}[1]{\nabla\cdot{{#1}}}

\def\be{\begin{equation}}
\def\en{\end{equation}}    

\newcommand{\bi}[1]{\mbox{\boldmath$#1$}}
\newcommand{\av}[1]{\langle{#1}\rangle}
\def\p{\partial}
\def\bea{\begin{eqnarray}}
\def\ena{\end{eqnarray}}

\def\gs{> \kern -12pt \lower 5pt \hbox{$\displaystyle{\sim}$}}
\def\ls{< \kern -12pt \lower 5pt \hbox{$\displaystyle{\sim}$}}

\title{Phase field model of
solid-liquid and 
liquid-liquid phase transitions 
in flow and elastic fields in one-component systems}
\shorttitle{Phase field model 
with hydrodynamics and elasticity }

\author{Kyohei Takae  and Akira Onuki}
\shortauthor{Takae  and  Onuki}

\institute{                    
Department of Physics, Kyoto University, 
Kyoto 606-8502, Japan
}

\abstract{We construct  a phase field model 
including hydrodynamics and elasticity 
in  one-component systems. 
It can be used to  investigate  solid-liquid 
and liquid-liquid phase transitions.  Upon first-order phase transition, 
a velocity field is 
induced around interfaces  
in the presence of a density difference 
between the two phases  even without applied 
shear flow.   As applications,  we present simulation results 
on  two cases of melting,  where 
a solid domain is placed 
on a heated wall in one case  and is suspended 
in  a warmer liquid under shear flow in the other.  
We find that the solid domain 
moves or rotates as a whole 
due to elasticity, releasing  latent heat. 
We also examine the liquid-liquid phase transition of 
a highly viscous  domain into  
a less viscous liquid on a heated wall, where an 
inhomogeneous  velocity field   is induced 
within a projected part of the domain. 
In these phase transitions, 
the interface temperature 
is nearly equal to the  coexisting 
temperature $T_{\rm cx}(p)$  away from the heated wall 
 in the presence of heat flow 
in the surrounding liquid. 
}

\begin{document}

\maketitle


\section{Introduction}

In solid-liquid phase transitions,  
 various nonequilibrium 
patterns have been 
observed \cite{Langer,Glicksman}. 
To reproduce such patterns numerically,   
phase field models 
have been  used extensively  \cite{Kobayashi,Karma}, where 
  a space-time dependent 
phase field $\phi({\bi r},t)$ 
 takes different values in  solid and liquid 
varying  smoothly across diffuse interfaces. 
A merit of this approach is 
that any surface boundary conditions need not 
be  imposed explicitly 
in simulations. Most theories of crystal growth 
 have assumed that 
the dynamics of first-order phase transition is  
governed by diffusion of  heat and/or composition. 
Though some attempts have  been made  
to include the velocity field or  convection into theory  
\cite{Amberg,Karma1999,Anderson,Goldenfeld}, 
 understanding of hydrodynamic  effects
during  solidification or melting still remains 
inadequate. 
Moreover, phase field calculations were performed 
on the surface instability 
in epitaxial film growth 
\cite{Grant,Misbah}, where 
elastic effects are crucial. 
It is worth noting that 
 various phase field models 
have been used to investigate phase ordering 
in  solid-solid phase transitions \cite{Onukibook}.

We also  mention  phase transitions in one component fluids 
between two liquid phases with 
 different microscopic structures 
\cite{Katayama,Tanaka2}.
In particular, Tanaka's group \cite{Tanaka2} 
 performed experiments of 
the phase ordering dynamics 
at   liquid-liquid  phase 
 transitions. To interpret their data, they   introduced 
 a nonconserved order parameter 
representing microscopic  structural order. 
From our viewpoint,  
we should develop a phase field model of liquid-liquid phase 
 transitions,  
where the structural order parameter, denoted by 
the same notation  $\phi$,  
is coupled to   the hydrodynamic variables.

One of the present authors has 
developed 
the dynamic van der Waals theory
for  one-component fluids  
 \cite{Onuki_vander,Teshi}.
It is a phase field  model of fluids 
 based on  the van der Waals theory. 
It can  treat   evaporation and condensation 
with inhomogeneous temperature 
accounting for latent heat. 
The aim of this letter is to 
present a phase field model 
including  both hydrodynamics and elasticity 
on the basis of well-defined thermodynamics.   
Our model  should  thus be applicable to 
 solid-liquid and liquid-liquid phase transitions. 
We will treat  one-component systems only 
for simplicity.  No gravity will be assumed.

\section{Phase-field  model}

We discuss thermodynamics 
and  dynamics  
for the phase field $\phi$   and the hydrodynamic variables.  
In the bulk region   $\phi$ is  equal to  
$0$ in liquid (or liquid phase I) and to $1$ in 
solid (or liquid phase II). 

At the starting point, 
we introduce an entropy density including a 
gradient contribution  as
\cite{Onuki_vander}
\be
\hat{S} =S(n,e,\phi)-\frac{1}{2} C |\nabla\phi |^2 ,
\en 
where $S(n,e,\phi)$ is a function of   the  number density 
$n$, the internal energy density $e$, and $\phi$. 
In this work 
 $C$ is a positive constant (which may depend on $\phi$ 
 more generally).
The temperature $T$ and the chemical potential $\mu$ 
are defined by $1/T= \p S/\p e$ and $\mu/T= -\p S/\p n$. 
The derivative with respect to $\phi$ is written as 
$\p S/\p \phi= -\Gamma/T$. The differential form of $S$ reads 
\be 
dS=(de-\mu dn -\Gamma d\phi)/T.
\en   
Neglecting  the gradient energy density \cite{Onuki_vander}, we 
assume the total energy density in the form,  
\be 
e_{\rm T}= e  +  \rho |{\bi v}|^2/2+ G(\phi)\Phi(e_2,e_3), 
\en 
where $\rho=mn$ is the mass density  with 
$m$ being  the molecular mass, $\bi v$ is the velocity field, 
and the last term is the elastic energy density. 
The shear modulus  $G(\phi)$ is 
zero for $\phi=0$ (in liquid) 
and is a positive constant for $\phi=1$ (in solid) 
for solid-liquid transitions, 
while $G=0$ for liquid-liquid transitions. 
 Here we suppose two dimensions, where 
anisotropic elastic 
strains are $e_2$ and $ e_3$. 
For small elastic deformations in solid, 
we have $e_2= \nabla_x u_x-\nabla_y u_y$,  
$e_3= \nabla_x  u_y+\nabla_y u_x$, and $\Phi=(e_2^2+e_3^2)/2$ 
 in terms of  the displacement field ${\bi u}= (u_x,u_y)$. 
Hereafter $\nabla_x=\p/\p x$ and $\nabla_y=\p/\p y$.
 In  the simulation in this work, 
 the amplitudes of $e_2$ and $e_3$ remain very small ($<10^{-3}$). 
However, the linear elasticity does not hold for large strains  
and a simple form of $\Phi$ applicable in the nonlinear regime 
is given by the periodic function \cite{Onuki-plastic} 
\be
\Phi=  \frac{1}{6\pi^2} 
\bigg [ 3-  \cos(2\pi e_2) - \cos(2\pi e_+)-
 \cos(2\pi e_-)
 \bigg ] 
\en  
where $e_{\pm}=(\sqrt{3}e_3\pm e_2)/2$. 
If we set $e_3 = e \cos\chi$ and 
$e_2 = e \sin\chi$, 
we have $\Phi = e^2/2 -{\pi^2}e^4/8
-{\pi^4} e^6 \cos (6\chi)/720+\cdots$. 
If we rotate  the reference 
frame by angle $\theta$, 
$\chi$ is shifted by 
$2\theta$. Thus  
$\Phi$ is highly isotropic for  small 
$e$ (less than $ 0.5$).

We follow  the principle of 
 nonnegative entropy production 
 to set up the dynamic equations. 
We introduce the generalized thermodynamic force 
associated with $\phi$ as 
\be
\hat{\Gamma}=  \Gamma  -TC \nabla^2\phi 
+  G'  \Phi, 
\en
where $G'=\p G/\p \phi$. The dynamic equation of $\phi$  is then  
\be
\frac{\p \phi}{\p t}= -{\bi{v}}\cdot\nabla\phi  -\Lambda \hat{\Gamma},
\en 
where   
$\Lambda$ is the kinetic coefficient. 
The   hydrodynamic equations are of the usual  forms,    
\bea
&&\hspace{-1cm}\pp{\rho}{t}=- \di{({\rho\bi{v}})}, \\
&&\hspace{-1cm}\pp{\rho\bi{v}}{t}=- \di{(\rho\bi{v}\bi{v}
+\ten{\Pi}-\ten{\sigma})},\\
&&\hspace{-1cm}\pp{e_T}{t}=-\di{[e_T\bi{v}+(\ten{\Pi}-\ten{\sigma})\cdot\bi{v}
-\lambda \nabla T]}. 
\ena  
The reversible stress  tensor $\ten{\Pi}=\{\Pi_{ij}\}$ 
contains the gradient and elastic parts  as  
\be
\Pi_{ij}=(p-G\Phi-\frac{1}{2}TC|\nabla\phi|^2)\delta_{ij}
+TC\nabla_i \phi \nabla_j \phi-G\epsilon_{ij}.
\en
where $p 
= n\mu-e +ST$ is the pressure. 
In two dimensions, the tensor $\epsilon_{ij}$ is given by 
by $\epsilon_{xx}= -\epsilon_{yy}= \p \Phi/\p e_2$ 
and $\epsilon_{xy}=\epsilon_{yx}=\p \Phi/\p e_3$. 
The  viscous stress tensor 
 $\ten{\sigma}=
 \{\sigma_{ij}\}= \{\eta (\nabla_i v_j +\nabla_j v_i)+ 
(\eta_B-2\eta/d) \delta_{ij}\nabla\cdot{\bi v}\}$ 
is written in terms of 
the shear viscosity $\eta $ and the bulk viscosity 
$\eta_B$ (in  $d$ dimensions). The strains $e_2$ and $e_3$ obey 
\bea 
&&\hspace{-1cm}\pp{e_2}{t}=-\bi{v}\cdot\nabla{e_2}+ \nabla_x v_x-\nabla_y v_y,\\&&\hspace{-1cm}\pp{e_3}{t}=-\bi{v}\cdot\nabla{e_3}+ 
\nabla_x v_y+\nabla_y v_x.
\ena 
With  these equations, the entropy density   $\hat{S}$  obeys  
\bea
\pp{}{t} \hat{S} &=&
-\di{\bigg[\hat{S}\bi{v}-\Lambda \hat{\Gamma}C\nabla\phi
-\frac{\lambda}{T}\nabla T\bigg]}\nonumber\\ 
&&+({\dot{\epsilon}_\theta+\dot{\epsilon}_{v} 
+\dot{\epsilon}_\phi})/{T},
\ena
where $\dot{\epsilon}_{v}$, 
$\dot{\epsilon}_\theta$, and 
 $\dot{\epsilon}_\phi$ 
are the heat production rates, 
\be
\dot{\epsilon}_\theta= \frac{\lambda}{T} (\nabla T)^2, \quad  
\dot{\epsilon}_v =
\sum_{ij}\sigma_{ij}\nabla_i v_j,\quad  
\dot{\epsilon}_\phi=\Lambda \hat{\Gamma}^2 .
\en

\section{Model entropy}

We suppose 
a reference equilibrium state 
at $T=T_0$ and $p=p_0=p_{\rm cx}(T_0)$, 
where   liquid and solid coexist macroscopically 
and  the chemical potential takes a common 
value $\mu=\mu_0=\mu_{\rm cx}(T_0)$.  
The quantities in the reference liquid state will 
be denoted with the subscript 
$0\ell$, while those 
in the reference solid state with the subscript 
$0s$. The number and energy  
densities in  the reference 
liquid (solid)  are written as 
 $n_{0\ell}$ and $e_{0\ell}$ 
($n_{0s}$ and $e_{0s}$) , respectively.  
The number and energy density 
 deviations are defined as those  from 
the reference liquid values  as  
$\delta n=n-n_{0\ell}$ and $\delta e=e-e_{0\ell}.
$

We propose to use 
a simple expression for 
 the  entropy density $S$ in Eq.(2.2). 
It contains terms up to second orders 
in $\delta n$ and $\delta e$ as 
\be
S=S_{0\ell}+\frac{\delta e-\mu_0 \delta n}{T_0}- 
C_V^0\frac{\tau^2}{2}  -\frac{\zeta^2}{2T_0K_T^0}-W(\phi).
\en
Here $S_{0\ell}$, $C_V^0$, and 
$K_T^0$ are  the entropy density, 
the constant-volume heat capacity  
per unit volume, 
and the isothermal 
compressibility  
in the reference liquid  state, respectively. 
We define   dimensionless variables $\tau$ and $\zeta$ by 
\bea
\tau &=&(\delta e- \beta_0 
 \delta n )/T_0C_V^0  - a_1\theta(\phi), \\
\zeta &=& \delta n/n_{0\ell}  -a_2\theta(\phi),
\ena
where  $\beta_0 =(\p e/\p n)_T$ is 
the derivative  in the reference liquid   
state. We define 
 $W(\phi)$ in $S$ and 
 $\theta(\phi)$ in $\tau$ and $\zeta$  
 as 
\be
W(\phi)=\frac{1}{2} A\phi^2(1-\phi)^2,
\quad \theta(\phi)=\phi^2(3-2\phi),
\en 
where $A$ is a positive 
constant. Then $\theta(0)=0$ and $\theta(1)=1$.
The  coefficients $a_1$ and $a_2$ in $\tau$ and $\zeta$ 
are related to the latent heat and  the density difference 
between liquid and solid, as will be shown below. 
For liquid 
with $\phi=0$, Eq.(15)  
is a well-known  expansion form 
up to second order deviations  \cite{Onukibook}. 
Adding the terms proportional to 
$\theta(\phi)$ in $\tau$ and $\zeta$, 
 we  use it   even for 
solid.

As functions of $\tau$, $\zeta$, and $\phi$, 
$T$, $\mu$, and $\Gamma$  are written as 
\bea
T &=& T_0/(1-\tau), \\
{\mu}  &=&  [{\mu_0} 
-{\beta_0\tau}
+{\zeta}/{n_{0\ell}K_T^0}]/(1-\tau), \\
{\Gamma}  &=& T\frac{d}{d\phi} [W(\phi) - n_{0\ell}h \theta(\phi)],  
\ena
where $h$ is the linear combination of $\tau$ and $\zeta$ defined by  
\be 
n_{0\ell}h=
(a_1 C_V^0) \tau +({a_2}/{k_BT_0K_T^0}) \zeta.
\en 
The pressure $p=n\mu+TS-e$ is then of the form,  
\be
\frac{p-p_0}{T} = \frac{\zeta+
T_0\alpha_p^0\tau}{T_0K_T^0}
+  C_V^0 \frac{\tau^2}{2} +\frac{\zeta^2}{2T_0K_T^0} -W+n_{0\ell}h\theta , 
\en 
where $\alpha_p^0= - (\p n/\p T)_p/n$ is 
the thermal expansion coefficient in the reference liquid state.

We consider equilibrium two-phase coexistence 
 without anisotropic strains.  
From Eqs.(19) and (20) 
$\tau$ and $\zeta$  take common values 
in the two phases, while Eqs.(21) and (23)  
yield  $h=0$.    
The liquid  phase $\phi=0$ is preferred  for 
$h<0$  and the solid   phase $\phi=1$ 
 for $h>0$.  Thus $h$ represents 
 the distance from the coexistence line 
 in the phase diagram.
From Eq.(17) the  density difference 
 between the two phases is  
given by 
\be 
\Delta n=n_s-n_\ell=a_2 n_{0\ell}.
\en  
The latent heat  per particle is $q=T(s_\ell-s_s)$, where 
 $s=S/n$. Its value 
in the reference state  ($\tau=0$) 
is written as   
\be 
q_0= B_1T_0 a_2/n_{0\ell}(1+a_2),  
\en
where $B_1= \alpha_p^0/K_T^0- C_V^0 a_1/a_2$ 
is equal to the derivative $(\p p/\p T)_{\rm cx}$ 
along the coexistence line for $\tau=0$. 
If $h=0$, we confirm the 
Clapeyron-Clausius relation $q= T(\p p/\p T)_{\rm cx} 
\Delta n/n_\ell n_s$ for any $\tau$. 
On the other hand, the  equation for a planar interface  
$\phi=\phi_{int}(z)$  becomes 
$W(\phi)= C |d\phi/dz|^2$, 
which is solved to give 
$ 
\phi_{int}(z)= 1/(1+e^{z/\xi}).  
$ 
The interface thickness 
is $\xi=(C/A)^{1/2}$ 
and the surface tension is   
$
\gamma=  TC/6\xi.
$

\begin{figure}[htbp]
\begin{center}
\includegraphics[width=230pt]{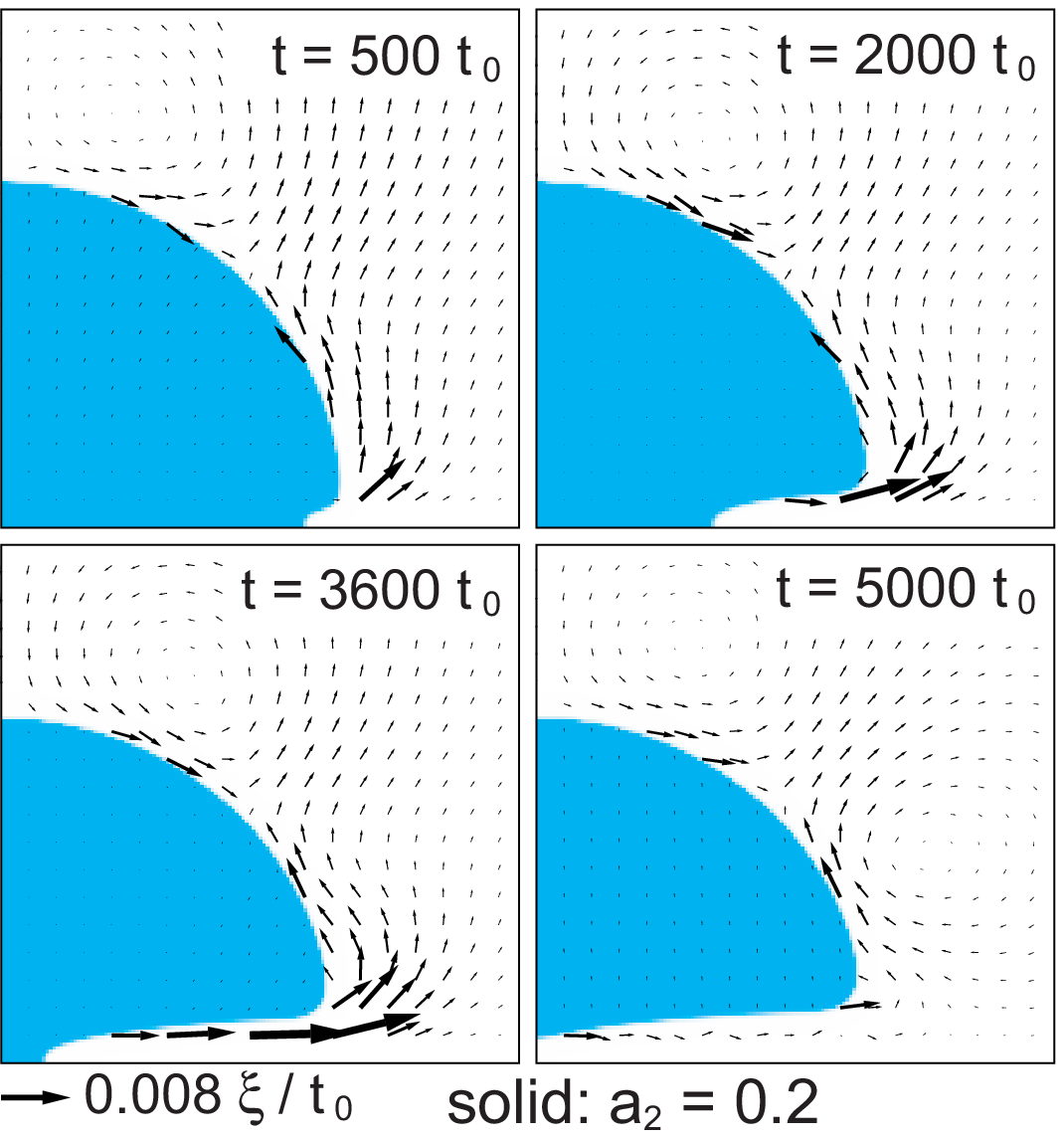}
\includegraphics[width=230pt]{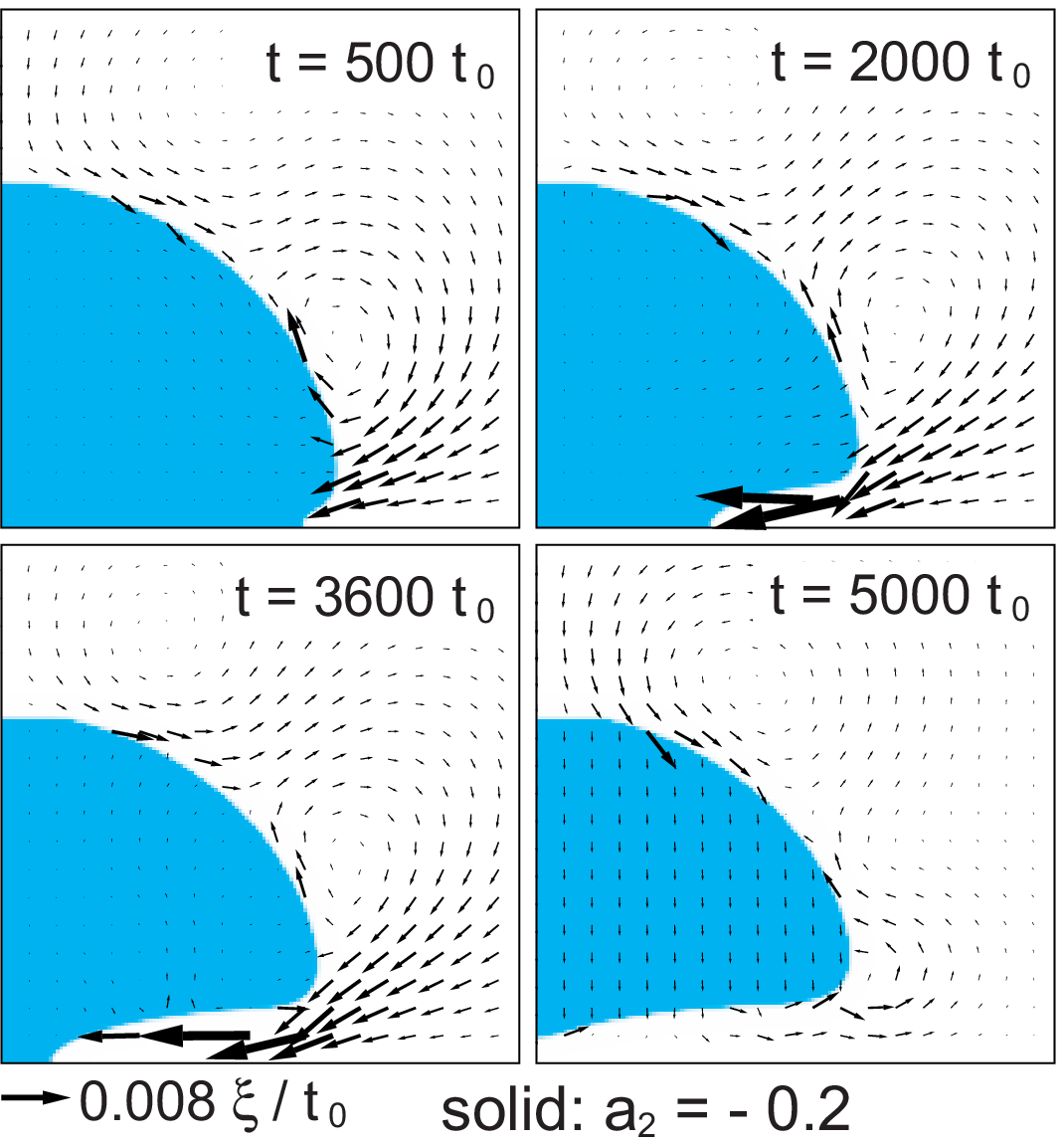}
\caption{Solid domain at four times 
on a heated wall  for $a_2=0.2$ (upper  four plates) 
and for $a_2=-0.2$ (lower four plates) 
in the region  $1/2<x/L<7/8$ and  $0<y/H<3/4$. 
Arrows  represent  the velocity field. 
Those  below the panels 
correspond to $0.008\xi/t_0$.  Melting mostly takes place  
close to the heated wall, where the velocity is from 
solid to liquid for $a_2=0.2$ and from 
 liquid to solid for $a_2= -0.2$. 
}
\end{center}
\end{figure}

\begin{figure}[htbp]
\begin{center}
\includegraphics[width=230pt]{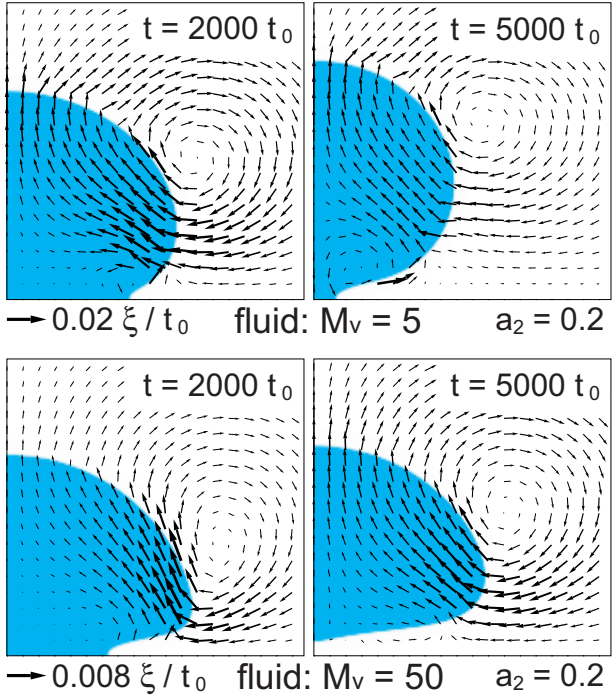}
\caption{
Droplet at liquid-liquid transition with $G=0$ 
at $t/t_0=2000$ and $5000$ on a heated wall 
in the region  $1/2<x/L<7/8$ and  $0<y/H<3/4$. 
The viscosity ratio $M_v$ is 
5 (top) and 50 (bottom).
A typical velocity is $0.02\xi/t_0$ 
for $M_v=5$ and $0.008\xi/t_0$ 
for $M_v=50$.  A circulating velocity field is marked.
See Fig.5 for the velocity along the interface.}
\end{center}
\end{figure}

\section{Numerical Method}

In two examples to follow, we integrated 
Eqs. (7), (8), and (11)-(13)
in two dimensions on a $800\times 400$ lattice. 
We use the entropy equation (13) 
instead of the energy equation (9) to suppress 
the so-called parasitic flow \cite{Teshi}, which is 
an artificial flow around an interface \cite{Jamet,Daru}.     
The simulation 
 mesh length is $\Delta x=\xi= (C/A)^{1/2}$. Here 
 we set $A= 0.4n_{0\ell}k_B$, so $C=0.4n_{0\ell}k_B\xi^2$. 
The horizontal and vertical lengths of the 
system are then  $L=800\xi$ and $H=400\xi$, respectively. 
We imposed the periodic boundary condition  
along the  horizontal $x$ axis and 
 the no-slip condition ${\bi v}={\bi 0}$ 
 at $y=0$ and $H$. In addition, 
 neglecting  the surface entropy and energy, 
we set  $\p \phi/\p z =0$ at $y=0$ and $H$.  

\begin{figure}[htbp]
\begin{center}
\includegraphics[width=250pt]{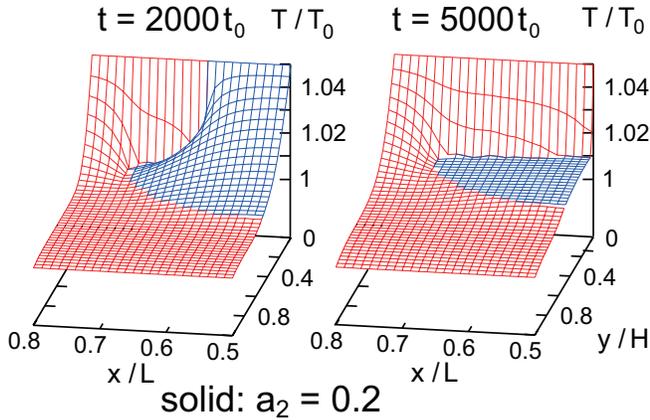}
\caption{Temperature in the region $0.5<x/L<0.8$ 
and $0<y/H<1$  for the solid case 
with $a_2=0.2$. See the corresponding domain shapes 
in Fig.1. At $t=2000t_0$ (left), heat from 
the wall is used to  melt the solid 
  near  the contact point. At $t=5000t_0$ (right), 
  the domain is detached from the wall and its 
  interface temperature becomes a constant. 
}
\end{center}
\end{figure}

\begin{figure}[htbp]
\begin{center}
\includegraphics[width=251pt]{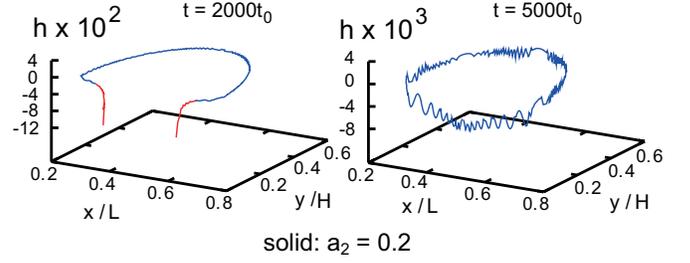}
\caption{$h$  in Eq.(22) 
of a melting solid  along  the interface, which  
is multiplied by $100$ at $t=2000t_0$ 
(left) and by $1000$ at $t=5000t_0$ (right). 
It represents   the distance from the coexistence 
curve, where $h>0$  
in the body part (blue)  and $h<0$  in the constricted 
part (red).   
}
\end{center}
\end{figure}

The thermodynamic quantities  are given by 
$C_p^0=1.1 C_V^0= 9.17 n_{0\ell}k_B$, 
$K_T^0=0.064/n_{0\ell }k_BT_0$, and 
$\alpha_p^0= 0.20/T_0$, which are consistent 
with the thermodynamic identity 
$T\alpha_p^2 = (C_p-C_v)K_T$ of 
fluids\cite{Onukibook}.
The shear modulus is 
$
G= 10 n_{\ell0}k_BT_0 \phi^2 
$ for solid-liquid transitions. 
By  setting  $a_1=-0.27$,  
we examine two cases of $a_2=\Delta n/n_{0\ell}=0.2$ and $-0.2$.
The latent heat $q_0$ in Eq.(25) is  
 $2.5k_BT_0$  
for $a_2=0.2$ and  
 $1.9k_BT_0$   for $a_2=-0.2$. 

The viscosities are given by 
$
\eta(\phi) =\eta_B(\phi)
=\eta_\ell +(\eta_{s}-\eta_{\ell})\phi^2
$. Let $\nu_0=\eta_\ell/mn_{0\ell }$ be the 
liquid kinematic viscosity.  
We measure space and time in units of 
$\xi$ and 
\be 
t_0= 0.4 \xi^2/\nu_0.
\en 
The viscosity  ratio 
$M_v= \eta_s/\eta_\ell$ 
is  taken to be 10 or 50. 
The simulation time mesh $\Delta t$ is $0.005 t_0$  
for  $M_v= 10$ and $0.0025 t_0$  for $M_v= 50$. 
The thermal conductivity is 
$\lambda= 1.63 \nu_0 C_p^0 (1+3\phi^2)$, 
so the  Prandtl number is  $1/1.63$ in liquid. 
The kinetic coefficient for $\phi$ 
is $\Lambda=0.16/n_{0\ell}k_BT_0 t_0$. 
With these expressions, the reference 
temperature $T_0$ and density $n_{0\ell}$ 
do not appear in the scaled dynamic equations.

\section{A semispheric domain  
on a heated substrate}

First, we 
  placed  a solid (type II liquid) 
  semisphere with radius $R=200\xi$ 
on the substrate in   liquid (type I liquid), 
where $\phi=1$ inside the domain, $\phi=0$ outside it,  
and $\tau=\zeta=e_2=e_3=0$ in  the whole cell. Thus 
$T=T_0$ and $\delta n =n_{0\ell} a_2 \theta(\phi)$ were 
imposed. 
The boundary temperatures at $z=0$ and $H$ were 
fixed at $T_0$. Then small relaxations 
followed  near the interface 
in short times ($\sim t_0$). 
After an equilibration  time  of $100t_0$, 
we raised the bottom temperature  to $1.05T_0$. 
We set $t=0$ at this heating.

\begin{figure}[htbp]
\begin{center}
\includegraphics[width=180pt]{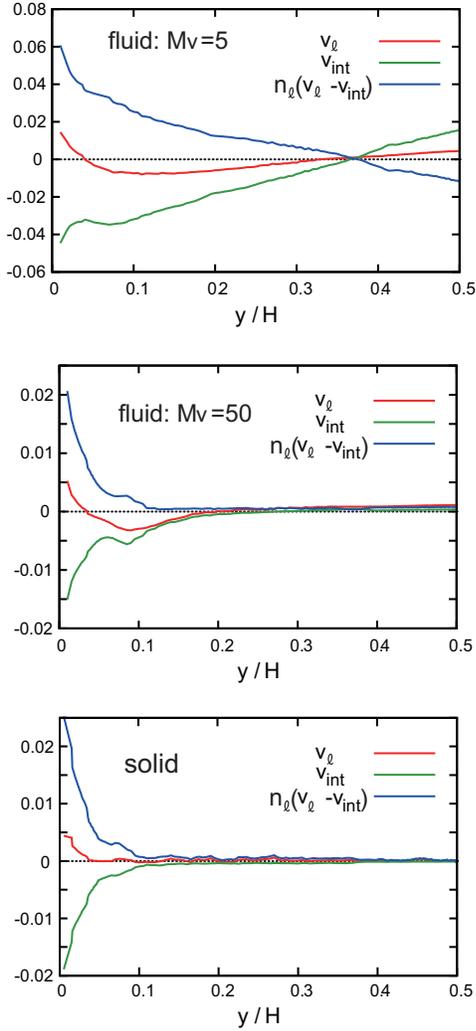}
\caption{ Liquid velocity $v_\ell$, interface velocity $v_{int}$, 
and melting flux $J= n_\ell(v_\ell-v_{int})$ in the normal 
direction  along the interface 
in units of $\xi/t_0$ or  $n_{0\ell}\xi/t_0$, where  $t=2000t_0$ and $a_2=0.2$.  
They are  for liquid-liquid 
transitions with $M_v=5$ (top) and 50 (middle) 
and  for a solid-liquid transition (bottom). 
Melting occurs mostly in the constricted 
part close to the heated wall. 
}
\end{center}
\end{figure}

In Fig.1, we show the  domain shapes  
and the surrounding velocity at four times. The solid 
  density is higher or lower than the liquid density  
depending on the sign of  $a_2$. 
The phase change takes place mostly 
near the heated wall. 
For $a_2=0.2$ the flow is from the solid to the liquid 
near the bottom in the upper plates, 
while for $a_2=-0.2 $  
the flow is in the reverse direction in the lower plates. 
The solid velocity is very small 
and is nearly uniform within the domain. 
The strains   
$e_2$ and $e_3$ remain   of order $10^{-3}$ and are 
well in the linear elasticity regime. 
Nevertheless, the resultant 
 shear stress ($\sim 10^{-2}n_{0\ell}k_BT_0$) 
 can realize   the solid body motion. 
In addition, the pressure $p$ in the liquid region 
gradually increases for $a_2>0$ 
and decreases for $a_2<0$ due to the density 
difference. In our examples the deviation 
 $p-p_0$ in the liquid is  of order 
$\pm 0.15n_{0\ell}k_BT_0$ for $a_2=\pm 0.2$ 
at $t=10^4t_0$.  
(Here the resultant adiabatic temperature 
change is very small since $(\p T/\p p)_s \sim 
0.03/ n_{0\ell}k_B$.)

Figure 2 gives  
the  profiles of a droplet 
in a less viscous liquid at 
 liquid-liquid transitions   with $G=0$. 
 The velocity field 
decreases with  increasing the viscosity ratio 
$M_v=\eta_s/\eta_\ell$, but it still remains  
noticeable in the projected part of the domain 
even for $M_v=50$. Here an interface motion 
induces  a fluid motion with a small velocity gradient 
within a highly viscous droplet. 
However,  the domain shapes 
 in the two cases in Figs.1 and 2 
 are very similar.

In Fig.3, the temperature is displayed 
for the solid case with $a_2=0.2$ before and after 
detachment of the   domain. 
Similar profiles were also obtained 
for the case of liquid-liquid transitions. 
A steep  gradient 
can be seen near the  heated wall, but the interface temperature 
is nearly flat at the melting 
temperature $T_{\rm cx}(p)$ 
far from the wall. Thus the interface  is divided into 
 the  body part far from the wall and 
the constricted part close to the wall. 
In this case  the thermal diffusion 
length $(D_Tt)^{1/2}\sim (t/t_0)^{1/2}\xi$ 
is still shorter than the cell height $H=400\xi$, 
resulting in  a small temperature gradient 
in the upper region. 
In Fig.4, we show that  
$h$ in Eq.(22) is very small  on the interface. 
Here $h \cong -2.5[\tau - (\p T/\p p)_{\rm cx} 
(p-p_0)/T_c]$ away from the interface (or for $\phi= 
0$ or 1 in Eq.(23)). 
In the constricted part,  
the gradient of $h$ is $-10\%$ of that of 
$\tau$.

In Fig.5, we plot the liquid velocity 
$v_\ell = {\bi \nu}\cdot {\bi v}$,  
the interface velocity $v_{int} = 
{\bi \nu}\cdot {\bi v}_{int}$, and 
the melting  flux $J= n_\ell(v_\ell-v_{int})$ 
through the interface, where $n_\ell$ is the liquid density. 
These are the quantities 
close to the interface in the normal direction 
$\bi \nu$.    
Indeed, melting mostly occurs 
in the constricted part  for large $M_v$ 
and for the solid case. We also notice that 
$|v_{int}|$ is considerably larger than 
$|v_\ell|$, which is obviously due to 
the small size of  $a_2$.

\section{A solid domain in a sheared warmer liquid}
\begin{figure}[htbp]
\begin{center}
\includegraphics[width=160pt]{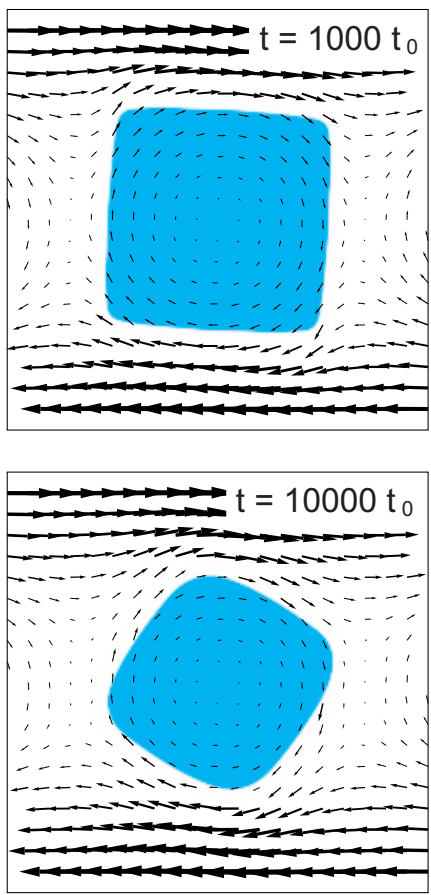}
\caption{A  square-shaped 
solid domain in a warmer liquid 
under  shear flow with $\dot{\gamma}=
2\times 10^{-4}t_0^{-1}$ at $\dot{\gamma} t= 0.2$ (top) 
and 2 (bottom). 
It gradually melts and 
rotates as a solid body. Melted particles 
are convected to cool the surrounding liquid. 
}
\end{center}
\end{figure}
\begin{figure}[htbp]
\begin{center}
\includegraphics[width=200pt]{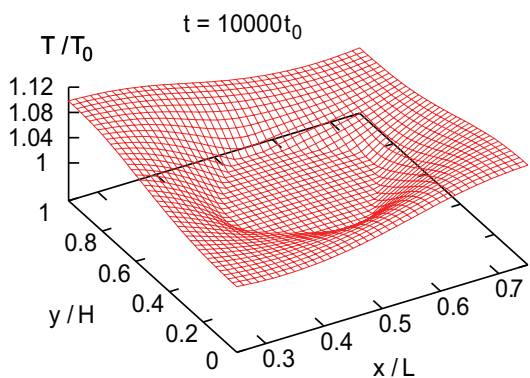}
\caption{Temperature profile 
around a melting solid in shear flow 
at $\dot{\gamma} t= 2$,  corresponding to the 
lower panel of Fig.6. It is 
lowest in the directions making  angles of 
 $\pi/4$ and $5\pi/4$ with respect to the $x$ axis due to convection 
 of cooler melted particles. 
}
\end{center}
\end{figure}

Next,  at $t=0$, 
we placed a 240$\times$240 solid 
square with $a_2=0.2$ in a liquid. 
The temperature was initially    $0.95T_0$ 
in the solid and $ 1.1T_0$ in the liquid. 
The top and bottom plates were   insulating 
or $\p T/\p z=0$.                                     
By moving them we  applied a shear flow with rate 
$\dot{\gamma}=2\times 10^{-4}t_0^{-1}$ for  $t>0$. 
The liquid  should be 
cooled upon melting.

Figure 6 displays the 
 solid shapes at $t=10^3t_0=0.2/\dot{\gamma}$ 
 and $10^4t_0=2/\dot{\gamma}$ 
 in the middle region $1/4<x/L <3/4$ and $0<y/H<1$, 
 where the solid  is rotating and melting.  
The  final solid area in Fig.6 is   
$70\%$ of the initial area. 
Figure 7 gives the temperature profile at 
$t=10^4t_0$. It  demonstrates 
a considerable cooling of the liquid particularly 
in the regions  where the melted liquid has been  convected.  
The  temperature is 
homogeneous with $h\cong 0$  within the solid domain. 
The average temperature $\av{T}=
\int_0^L dx\int_0^H dy T({\bi r},t)/LH$ 
is $1.078T_0$  at $t=0$, 
$1.076T_0$  at $t=10^3t_0$, and 
$1.063T_0$  at $t=10^4t_0$. 
To explain the cooling by $0.015T_0$  at $t=10^4t_0$, 
we multiply 
the number of melted particles ($\sim 4000$) 
by the latent heat per particle $q_0$ in Eq.(25) 
and divide it by the total specific heat $LH C_v^0$  
to obtain a temperature decrease $\sim 0.015T_0$ in accord 
with the data of $\av{T}$.

\section{Summary}

 In our theory, the  entropy   depends on   the phase field $\phi$ 
 and contains a gradient part ($\propto |\nabla\phi|^2$),  
while the total  energy  consists of 
the usual internal energy, 
the kinetic energy, and the elastic energy. 
The elasticity is introduced  using 
 strain fields $e_2$ and $e_3$ in two dimensions, 
which represent anisotropic elastic deformations.  
Starting with these quantities and using the principle 
of nonnegative entropy production, we have constructed 
dynamic equations for the phase field, 
the hydrodynamic variables, and the strains. They can 
  describe phase transition dynamics accounting for 
  the  hydrodynamic and elastic effects. 
If the strains are absent or the 
shear modulus vanishes, we 
obtain the dynamic equations applicable to liquid-liquid  phase transitions. 
We have also proposed to use the entropy density 
containing  second-order deviations 
of the number and internal 
energy densities  from a reference 
two-phase state. 
We have solved the dynamic  equations 
to examine  two melting phenomena in two dimensions. 
In these cases  small strains 
of order $10^{-3}$ have  produced    
rigid  body motions of a solid domain.

There can be a number of 
 problems to be  studied in our scheme 
such as 
 dendrite formation in flow, spinodal decomposition, 
epitaxial growth, and recrystallization.
In particular, it is  of great interest to investigate 
kinetics with large strains 
or even with dislocations. 
In liquid-liquid transitions, 
we should further investigate the effects of 
latent heat and shear flow in phase ordering.

\acknowledgments
This work was supported by KAKENHI (Grant-in-Aid for Scientific Research) on Priority Area gSoft Matter Physicsh from the Ministry of Education, Culture, Sports, Science and Technology of Japan.


\end{document}